\documentstyle[12pt]{article}
\newcommand{\comma}{\;\; ,}
\newcommand{\period}{\;\; .}
\newcommand{\eq}{\; = \;}
\newcommand{\sep}{\;\;\; , \;\;\;}
\newcommand{\be}{\begin{equation}}
\newcommand{\oW}{\overline{W}}
\newcommand{\oV}{\overline{V}}

\newcommand{\bd}{\begin{displaymath}}

\newcommand{\ee}{\end{equation}}
\newcommand{\ed}{\end{displaymath}}
\newcommand{\ba}{\begin{eqnarray}}
\newcommand{\ea}{\end{eqnarray}}

\newcommand{\half}{\textstyle \frac{1}{2}}

\def\picture #1 by #2 (#3){
  \vbox to #2{
    \hrule width #1 height 0pt depth 0pt
    \vfill
    \special{picture #3}}}
\def\scaledpicture #1 by #2 (#3 scaled #4){{
  \dimen0=#1 \dimen1=#2
  \divide\dimen0 by 1000 \multiply\dimen0 by #4
  \divide\dimen1 by 1000 \multiply\dimen1 by #4
  \picture \dimen0 by \dimen1 (#3 scaled #4)}}

\title{ Functional relations for the order parameters \\
of the chiral Potts model: low-temperature expansions}

\author{ R.J. Baxter\\
{\protect \small Theoretical Physics, I.A.S. and School of Mathematical
Sciences}\\
{\protect \small  The Australian National University,
 Canberra, A.C.T. 0200, Australia} 
}

\begin{document}

\maketitle

\abstract{This is the third in a series of papers in which we set up and 
discuss the functional relations for the ``split rapidity line'' correlation function
in the $N$--state chiral Potts model. The order parameters of the model can be 
obtained from this function. Here we consider the case $N = 3$ and 
write the equations explicitly in terms of the hyperelliptic functions parametrization. 
We also present four-term low-temperature series expansions, which we hope will cast light
on the analyticity properties needed to solve the relations. The problem remains unsolved, 
but we hope that this will prove to be a step in the right direction.\\

\noindent {\protect \small PACS: 05.50.+q, 02.30.Gp} {\protect \small {\it Keywords:} 
Lattice models, chiral Potts model, 
order parameters, hyperelliptic functions }

}

\section{Introduction}

In a previous paper \cite{fnlrelns98} we used the method pioneered by 
Jimbo, Miwa and Nakayashiki \cite{JMN93} to obtain functional relations for a 
generalized
one-spin correlation function for the planar solvable $N$--state chiral Potts model.
The correct solution of this would give the spontaneous magnetization - a 
long-standing problem
which is all the more tantalizing as there is an elegant conjecture for the result
\cite{Howes83} -- \cite{HenkelLacki89}.

Unfortunately, the functional relations are really just symmetry and periocity conditions and 
have more than one solution.
One needs to supplement them with an appropriate analyticity requirement. For 
other models this is easy:
one goes to the elliptic (or trigonometric) parametrization that uniformizes 
the equations and
manifests the  rapidity-difference-property. There is then an obvious vertical 
strip in the complex plane of the
spectral parameter variable within which the function is analytic and Fourier 
transformable in the vertical direction. The equations can then be solved 
directly and uniquely
by taking such a transform, in much the same way as the inversion relations 
for the free energy can be solved \cite{Baxinv82}.

For the chiral Potts model with $N > 2$, there is no
such parametrization. In one sense one can uniformize the equations, but at the price of 
introducing hyperelliptic functions with more than one argument, the arguments being
related to one another in a complicated manner. We have discussed these functions
 (particularly for the case $N = 3$) earlier \cite{Kyoto91} -- \cite{CTMP}, and in
\cite{hypfns98} have obtained formulae for the coefficients in the 
functional relations. Here we write the 
relations in terms of this parametrization, focussing on the 
case $N = 3$. We discuss their properties and present some low-temperature expansions
of the functions, obtained by direct calculation from their definitions. We still have not
succeeded in obtaining the appropriate solution of the functional relations, but hope
that the data presented herein will assist in the search. Certainly the solution must
agree with the low-temperature expansions. 

\section{Functional relations}

The chiral Potts model is explained in \cite{fnlrelns98}. Spins live on the sites of a square lattice,
oriented diagonally. Each spin takes the values $0, \ldots , N-1$.
Adjacent spins $a, b$ interact with weight function $W_{pq}(a - b)$ on 
SW $\rightarrow$ NE edges, $\oW_{pq}(a - b)$ on SW $\rightarrow$ NE edges. The 
symbol $p$ denotes a point (a ``rapidity'') $a_p, b_p, c_p , d_p$ on the projective 
curve
\be \label{restrict1}
a_p^N  + k' \,b_p^N  \eq  k \, d_p^N \sep
k' \, a_p^N  + b_p^N  \eq  k \, c_p^N \comma  \ee
where $k, k'$ are  real constants (moduli) satisfying
\be k^2 + {k'}^2 \eq 1 \period \ee
The symbol $q$ denotes another such point and
\be \label{defWW}
W_{pq} (n)  \eq  \prod_{j=1}^n \frac{d_p b_q - a_p c_q \omega^j }
{b_p d_q - c_p a_q \omega^j } \sep
\oW_{pq} (n)  \eq  \prod_{j=1}^n \frac{\omega a_p d_q - d_p a_q \omega^j }
{c_p b_q - b_p c_q \omega^j } \period \ee

Operators $R$ and $M$ act on the rapidities, being defined by
\be
( a_{Rp} , b_{Rp} , c_{Rp} , d_{Rp} ) \eq 
(b_p, \omega a_p, d_p , c_p ) \comma \ee  
\be
( a_{Mp} , b_{Mp} , c_{Mp} , d_{Mp} ) \eq 
(\omega a_p, b_p , c_p , \omega d_p ) \period \ee 
They satisfy  $R M = M^{-1} R$ and  ensure that
\be \label{rotate}
\oW_{pq} (n) \eq W_{q,Rp}(n) \sep
W_{pq} (n) \eq \oW_{q,Rp}(-n) \period \ee

In \cite{fnlrelns98} we consider a square lattice with one spin deep inside it fixed
at some value $a$. The horizontal rapidity line immediately beneath this spin is 
broken, the left and right  segments having different rapidities $p, q$, respectively. 
The boundary spins are fixed to zero. Writing the corresponding partition 
function as $Z_{pq}(a)$, 
the probability that a free spin in this position has value $a$ is simply
\be \label{defFpq}
F_{pq}(a) \eq Z_{pq}(a)/[Z_{pq}(0) + \cdots + Z_{pq}(N-1)] \period \ee
In the limit of a large lattice, it is independent of the rapidities of all 
other lines (because of $Z$-invariance \cite{RJB78}). By definition, 
$F_{pq}(0) + \cdots + F_{pq}(N-1) = 1$.

In \cite{fnlrelns98} we show that the Yang-Baxter relations imply that 
$F_{pq}(a)$ satisfies the following functional 
relations:
\ba \label{Frelns}
 F_{Rp, Rq} (a)  \eq  F_{pq} (-a) & \sep &
F_{R q, p}(a)  \eq  F_{R p, q}(a) \comma \nonumber \\
F_{p, R q}(a) & \eq & \xi_{pq} \sum_{b=0}^{N-1} \,  \oW_{pq}(a - b) \,
 F_{q, Rp} (b)
\comma \\
F_{Mp, q} (a)  \eq   F_{p, M^{-1} q} (a) & \eq  & 
\eta_{pq} \, \omega^a F_{pq} (a)  \comma \nonumber \ea
where 
\be \xi_{pq} \eq \sum_{n=0}^{N-1} \; \oW_{pq} (n) \sep \eta_{pq} \eq \sum_{a=0}^{N-1} \omega^a \, F_{pq}(a) 
\period \ee

Let $\omega = \exp (2 \pi i /N)$. It is natural to introduce the discrete 
Fourier transforms (for $r$ any integer)
\ba  \label{deftF}
\tilde{F}_{pq} (r)  & \eq & \sum_{a=0}^{N-1} \omega^{r a} \, F_{pq} (a) \comma \nonumber \\
\oV_{pq} (r) & \eq & \sum_{a=0}^{N-1} \omega^{r a} \; \oW_{pq} (a) \period \ea
Then 
$\tilde{F}_{pq}(0) = \tilde{F}_{pq}(N) =1$ and the relations (\ref{Frelns}) become
\ba \label{Foureqns}
 \tilde{F}_{Rp, Rq} (r)  \eq  \tilde{F}_{pq} (N-r) & , &
\tilde{F}_{Rq, p} (r)  \eq  \tilde{F}_{Rp,q} (r) \comma \nonumber \\
 \tilde{F}_{p, Rq} (r) \! \! & = & \! \! \oV_{pq}(r) \, \tilde{F}_{q,Rp} (r) /\, \oV_{pq}(0) \comma  
\\
\tilde{F}_{Mp, q} (r) \eq  \tilde{F}_{p, M^{-1} q} (r)  \! \! & = & \! \!
 \tilde{F}_{pq} (r+1) / \tilde{F}_{pq}(1)
\period \nonumber \ea
Without loss of generality, in these equations we can take 
$r = 1,\ldots ,N-1$.

It can be convenient to work 
with the ratios:
\be \label{defG}
G_{pq}(r) \eq \tilde{F}_{pq}(r)/\tilde{F}_{pq}(r-1)  \ee
and with the product
\be \label{defL}
L_{pq} (r) \eq G_{pq}(r) G_{Rq, Rp}(r) \period \ee

In terms of $L_{pq}(r)$, the functional relations become, for 
$r = 1, \ldots , N$, with \\ $L_{pq} (N+1) = L_{pq} (1)$:
\ba \label{Lrelns}
L_{Rp,Rq}(r) & \eq & L_{qp}(r) \eq 1/L_{pq}(N\! -\! r\! + \! 1) \comma \nonumber \\
L_{p, Rq} (r) & \eq & \oV_{pq}(r) \,  L_{q,Rp} (r) / \oV_{pq}(r-1)  \comma \nonumber \\
L_{pq}(1) \! \! & \! \! \cdots \! \! & \! \! L_{pq}(N)   \eq  1 \comma \\
L_{Mp, q} (r) & \eq & L_{p, M^{-1} q} (r) \eq   L_{pq} (r+1) \period \nonumber \ea
It follows that 
\be
L_{R^{2N}p,q}(r) \eq L_{p, R^{2N}q}(r) \eq (-1)^{N-1} L_{pq}(r) \comma \ee
which is a periodicity (or anti-periodicity) property.

The function $F_{pp}(a)$ is the probability, for the regular square lattice,
that a central spin has value $a$. This should be independent of the rapidity $p$
\cite{RJB78}, so
$F_{pp}(a), \tilde{F}_{pp}(r)$, $G_{pp}(r), L_{pp}(r)$ should all be independent of $p$.
The order parameters are the expectation values ${\cal M}_r$ of $\omega^{r a}$ for 
$r = 1, \ldots , N-1$. From the above definitions of 
$\tilde{F}_{pq}(r), G_{pq}(r), L_{pq}(r)$, we see that
\be \label{emmr}
{\cal M}_r \eq  \tilde{F}_{pp}(r) \eq G_{pp}(1) \cdots G_{pp}(r) \eq \sqrt{
L_{pp}(1) \cdots L_{pp}(r) } \period \ee

The conjectured result for ${\cal M}_r$ is
 \be \label{conj}
{\cal M}_r \eq \langle \omega^{r a} \rangle \eq (1 - {k'}^{2} )^{r(N-r)/2N^2} 
\comma \ee
(eqn 3.13 of \cite{Howes83} , eqn. 1.20 of \cite{AMPT89} , eqn. 15
of \cite{HenkelLacki89}, $\beta$ and $\lambda$ therein being the $k'$ of
this paper. This $k'$ can be thought of as a temperature variable: small at low 
temperatures and increasing to one at criticality.

\section{Hyperelliptic function parametrization}
The hyperelliptic function parametrization is given in \cite{Kyoto91} and 
\cite{hypfns98}. There are $N-1$ constants $\rho_1 , \ldots , \rho_{N-1}$
(positive pure imaginary) which are defined solely by $N, k$ and $k'$, and satisfy 
$\rho_{\alpha} = \rho_{N - \alpha}$. For 
$\alpha, \beta = 1, \ldots , N-1$, define
\be \label{deftau}
\tau_{\alpha \beta} = \rho_{\alpha} + \rho_{\beta} - \rho_{|\alpha - \beta |}
 \comma \ee
taking $\rho_0 = \rho_N = 0$.
Let $s = \{ s_1 , \ldots , s_{N-1} \}$ be a set of variables. Then the
 hyperelliptic $\Theta$ function is defined by
\be \label{Theta}
\Theta(s) \eq \sum_m \exp \{ 2 \pi i \sum_{\alpha} m_{\alpha} s_{\alpha} + \pi i \sum_{\alpha}
\sum_{\beta} m_{\alpha} 
\tau_{\alpha \beta} m_{\beta} \} \comma \ee
the inner sums being over $\alpha, \beta = 1, \ldots , N-1$,
and the outer sum over all values of the integers $m = \{ m_1 , \ldots , m_{N-1} \}$.

We also define the constant sets $g, \rho$ by $g = \{ g_1, \ldots , g_{N-1} \}$,
$\rho = \{ \rho_1 , \ldots , \rho_{N-1} \}$, where $g_{\alpha} = \alpha / N$. From 
(\ref{deftau}) and (\ref{Theta}) the 
$\Theta$ function satisfies various quasi-periodicity and symmetry relations,
in particular
\be \label{Thsymm}
\Theta ( s) \eq \Theta(-s) \eq \Theta(\tilde{s}) \eq \Theta( s + N g ) \eq \exp \{ \pi i \sum_{\alpha} (2 s_{\alpha} + N 
\rho_{\alpha} ) \} \; \Theta ( s + N \rho ) \comma \ee
where $\tilde{s} = \{ s_{N-1} , s_{N-2} , \ldots , s_1 \}$ is the set$s$ in reverse order.

For each rapidity $p$ there is a set of such $s$--variables, which we write as $s_p$
(Greek, numerical or ``$N$'' suffixes refer to the position of an entry in the set, while 
lower-case Roman suffixes 
-- usually $p$ or $q$ -- refer to the rapidity dependence). We write the entry $\alpha$
of $s_p$ as $(s_p)_{\alpha}$. The entries $(s_p)_1, \ldots , (s_p)_{N-1}$ are not independent:
they are functions of the single rapidity variable $p$ and 
satisfy the $N-2$ relations (36) of \cite{Kyoto91}, or equivalently (10) of \cite{hypfns98}.

Set $t = \{ t_1 , \ldots , t_{N-1} \}$, where
\be 
t_{\alpha} \eq s_1 + \cdots + s_{\alpha} - \alpha (s_1 + \cdots + s_{N-1} )/N \period \ee
As with $s$, we write  $t_p$ for the set $t$
corresponding to the  rapidity $p$, and $(t_p)_{\alpha}$ for its entry $\alpha$. Then
\be \label{trp}
 (t_{Rp})_{\alpha} \eq (t_p)_{N - \alpha} + \half \rho_{\alpha} \comma \ee
\be \label{tr2p}
t_{R^2 p} \eq t_p + \rho \sep t_{Mp} \eq t_p - g \period \ee
(choosing $M$ so that $M^{-1}$ is the operator $M_1^{(1)} \cdots M_{N-1}^{(1)}$ of 
\cite{Kyoto91} and \cite{hypfns98}.)

The relation (\ref{trp}), together with the $s, \tilde{s}$ symmetry in (\ref{Thsymm}), 
implies that
\be \label{ThRR}
\Theta (t_{Rq} - t_{Rp} + m g + n \rho ) \eq \Theta (t_{q} - t_{p} - m g + n \rho ) 
 \ee
for all integers $m, n$.

From (\ref{defWW}) and (\ref{deftF}), 
\bd
\oV_{pq}(r) /\oV_{pq}(r-1) \eq (c_p b_q - \omega^r a_p d_q  )/
(b_p c_q - \omega^r d_p a_q )   \ed
so \be \label{ratV}
\oV_{p,Rq}(r) /\oV_{p,Rq}(r-1) \eq (\omega c_p a_q - \omega^r a_p c_q  )/
(b_p d_q - \omega^r d_p b_q ) \period  \ee
In eqn (26) of \cite{hypfns98} we give a formula for the RHS of (\ref{ratV}). This formula 
has not been proven for  $N > 3$, but we have tested it numerically and 
believe it should 
be proveable by Liouville's theorem, along the lines indicated in \cite{hypfns98}.
Using it, we obtain 
\be \label{VVpsipsi}
\oV_{p,Rq}(r)/\oV_{p,Rq}(r-1) \eq \psi_{p,R^2q}(r)/\psi_{pq}(r) \comma \ee
where
\be \label{defpsi}
\psi_{pq}(r) \eq \frac{\Theta (t_q - t_p + r g  )}
{\Theta [t_q - t_p + (r-1) g  ]} \period \ee

Replacing $q$ by $Rq$, the second line of equation (\ref{Lrelns}) is
(using the first equation)
\be \label{secondL}
 L_{p,R^2q}(r) \eq  \psi_{p,R^2q}(r) L_{pq}(r)/\psi_{pq}(r) \period \ee

\subsection*{A spurious solution}
  Obviously a solution of (\ref{secondL}) is
\be \label{spurious}
L_{pq} (r) \eq \psi_{pq}(r) \period \ee
Further, using (\ref{tr2p}) and (\ref{ThRR}), we find that it satisfies all the other 
equations (\ref{Lrelns}).
For $N = 2$, it is indeed the correct solution, but for general $N$ the 
formula (25) of \cite{hypfns98} (also unproven for $N > 3$, but believed to 
be ``proveable'') 
tells us that
\be
\psi_{pq}(r) \eq k^{(N-1)/N} \prod_{m=1}^{N-1} \;
\left\{ \frac{b_p d_q  - \omega^{j-m-1} d_p b_q }{c_p d_q - \omega^{j-m-1} d_p c_q}
\; \frac{c_p a_q  - \omega^{j-m-1} a_p c_q}{b_p a_q  - \omega^{j-m-1} a_p b_q} 
\right\}^{m/N}  \period \ee
This is precisely the wrong solution discussed in \cite{fnlrelns98}. It gives a 
result for ${\cal M}_r$ that is in error by a power $N/2$, and disagrees with 
low-temperature series expansions. Yet it is such a simple and elegant (in terms of 
our hyperelliptic functions)
solution of (\ref{Lrelns}).  This illustrates well the difficulty with functional
relations such as (\ref{Lrelns}): they do not by themselves define the function.
One has to incorporate the correct analyticity properties.

\section{The case $N = 3$}

If $N = 3$, then $\rho_1 = \rho_2 = \rho$, so if we define
\be x  \eq e^{2 \pi i \rho } \sep |x| < 1 \comma \ee
then
\be \Theta \{ s_1, s_2 \} \eq \Phi (e^{2 \pi i s_1}, e^{2 \pi i s_2} ) \comma \ee
where
\be \label{Phidefn}
\Phi(\alpha, \beta ) \eq \sum_{m,n} x^{m^2+m n+ n^2}
\alpha^m \beta^n \comma \ee
the sum being over all integers $m, n$. (Here we regard $x$ as a given constant and do 
not usually explicitly exhibit the dependence of functions on it.)  
The function $\Phi (\alpha, \beta)$
satisfies the symmetry and quasi-periodicity properties (34) --(36) of
\cite{hypfns98}, in particular
\be \label{Phisymm}
\Phi (\alpha, \beta) \eq \Phi (1/\alpha,1/\beta) \eq \Phi (\alpha, \alpha/\beta) \period \ee
 The nome $x$ is related to the modulus $k$ by \cite{Como93}
\be \label{kandx}
(k'/k)^2 \eq 27 \, x \;  \left[ Q(x^3)/ Q(x) \right] ^{12} \comma \ee
where
\be Q(x) \eq \prod_{n=1}^{\infty} (1-x^n) \period \ee
At low temperatures both $k$ and $x$ are small.

 If we define
\ba \alpha \eq & \alpha_{pq} & \eq \exp \{ 2 \pi i [(s_q)_1 - (s_p)_1 ] \} \comma \nonumber \\
 \beta \eq & \beta_{pq} & \eq \exp \{ 2 \pi i [(s_q)_2 - (s_p)_2 ] \} \comma 
\\
u \eq &  u_{pq} & \eq \exp \{ 2 \pi i [(s_q)_1-2 (s_q)_2 - (s_p)_1 + 2 (s_p)_2 ]/3
   \} \comma \nonumber  \ea
\be 
\rho (z) \eq \prod_{n=1}^{\infty}
(1 - x^{3n-2} z ) (1-x^{3n-1} / z)  \sep \phi(z) \eq \rho(z^{-1}) /\rho(z) \comma \ee
then \cite{Como93}
\be \label{aub}
\alpha \eq u^3 \beta^2 \comma \ee
\be \label{Wpq}
W_{pq}(1)  \eq  u^2 \beta \, \phi (\alpha ) \phi (\alpha /\beta ) \sep 
W_{pq}(2)  \eq  u   \beta \, \phi (\alpha ) \phi (\beta ) \period \ee 

For any function,  we can in principle eliminate the two degrees of 
freedom $p$ and $q$ in favour of the variables $\alpha, \beta$  (regarding $k, k', \rho, x$
as given constants). The resulting expressions will not necessarily be simple: they may be
multi-valued functions of $\alpha$ and $\beta$, but the simple form of (\ref{Wpq})
gives some encouragement that this may be a useful parametrization for our functions.
In this spirit, let us define a function $\psi(\alpha, \beta )$ such that
\be \label{F1}
F_{pq}(1) /F_{pq}(0) \eq x \, u^{-1} \beta^{-1} \,  \psi (\alpha, \beta ) \period \ee

We have expanded $\psi(\alpha, \beta )$ to third order in the low-temperature parameter
$x$, working directly from the definition (\ref{defFpq}) and performing finite-size 
lattice calculations.  The results are given below. We 
used a Fortran computer program, working with 
series in $x$ with double-precision numerical
coefficients for various given values of $\alpha$ and $\beta$. The method is not rigorous, 
depending both on observing that a given coefficient stabilized once 
the lattice was sufficiently large (the biggest lattice we considered 
was about eight by eight), and on
numerically fitting  the results to postulated expressions with unknown integer
coefficients. However, we believe the results to be correct. The key observation is that 
for $n > 0$ the coefficient of $x^n$ in the expansion is a Laurent polynomial in  
$\alpha$ and $\beta$, divided by 
$(\alpha - 1)^{2n-1}$. The coefficient of $x^0$ is one.

The variables  $\alpha_{pq}, \beta_{pq}, u_{pq}$ satisfy the relations
\bd \alpha_{Rp,Rq} \eq \alpha_{pq} \sep \beta_{Rp,Rq} \eq \alpha_{pq}/\beta_{pq} \sep
u_{Rp,Rq} \eq 1/u_{pq} \comma \ed
\bd \alpha_{qp} \eq 1/\alpha_{pq} \sep \beta_{qp} \eq 1/\beta_{pq} \sep
u_{qp} \eq 1/u_{pq} \comma \ed
\be \label{transab}
\alpha_{p,R^2 q} \eq x^2 \alpha_{pq} \sep \beta_{p,R^2 q} \eq x \beta_{pq} \sep
u_{p,R^2 q} \eq u_{pq} \comma \ee
\bd \alpha_{Mp,q} \eq \alpha_{p,M^{-1}q} \eq \alpha_{pq} \sep
  \beta_{Mp,q} \eq  \beta_{p,M^{-1}q} \eq \beta_{pq}  \comma \ed
\bd  u_{Mp,q} \eq  u_{p,M^{-1}q} \eq \omega^{-1}  \, u_{pq}  \period \ed
Using these, the first of the equations (\ref{Frelns}) is equivalent to
\be \label{F2}
F_{pq}(2) /F_{pq}(0) \eq x \, u^{-2} \beta^{-1} \,  \psi (\alpha, \alpha/\beta ) 
\period \ee

Using (\ref{transab}), (\ref{Wpq}) and (\ref{rotate}), the second and third of the relations
 (\ref{Frelns}) become
\be  \label{psireln1}
x \alpha \psi(x^{-2}/\alpha, x^{-1} \beta/\alpha ) \eq \psi(\alpha, \beta ) \comma \ee
\be  \label{psireln2}
\Delta (\alpha, \beta)  \, \psi(x^2/\alpha, x/\beta ) \eq \phi(\alpha ) \phi (\beta )
+ ( x/\alpha) \, \psi(\alpha, \alpha/\beta ) + (x/\beta)  \phi(\alpha ) 
\phi (\alpha /\beta ) \psi(\alpha, \beta ) \comma \ee
where
\be \label{defDelta}
\Delta  (\alpha, \beta) \eq 1 +  x \, \phi(\alpha ) 
\phi (\alpha /\beta )  \psi(\alpha, \alpha/\beta ) + x \, 
\phi (\alpha ) \phi (\beta ) \psi(\alpha, \beta ) \period \ee

The last two of the relations (\ref{Frelns}) (those 
involving $Mp$ and $M^{-1} q$) are automatically satisfied.

\subsection*{Series expansions for $\psi(\alpha, \beta )$}

Define 
\be \label{defvvbw}
v \eq \frac{\beta -1}{\beta (\alpha - 1)} \sep \overline{v} \eq 
\frac{\alpha - \beta }{\beta (\alpha - 1)} \sep  \mu \eq \beta v \overline{v}  \period \ee
Then, using the configuration of the rapidity lines $p$ and $q$ mentioned above, taking
$\alpha$ and $\beta$ to be of order one,  to third order in an expansion 
in powers of $x$ we find that:
\ba \label{psi1}
 && \psi(\alpha, \beta ) \eq 1 + x [ v \alpha  (\beta - 1) -2 + \alpha  ] + 
x^2   [
v \mu (\alpha^2-\beta)+v (1+4 \alpha+2 \alpha^2 - \nonumber \\ && 
2\beta -  5 \alpha \beta ) +
 13 - 6 \alpha ] +     x^3 [  v \mu^2 \alpha  (3 \beta- \alpha \beta - \alpha - \alpha^2) +
v \mu (2 \beta/\alpha + 5 \beta + \nonumber \\ & &  3 \alpha \beta + 
 1 - 2 \alpha - 8 \alpha^2 - \alpha^3 ) +
 v (-6 \beta /\alpha   +12 \beta + 28 \alpha \beta + \alpha^2 \beta 
 -3-21 \alpha -  \nonumber \\ & &  10\alpha^2-\alpha^3)
+ 5/\alpha -  71+38 \alpha + \alpha^2 ] + {\rm O } (x^4)
 \period  \ea

We also considered the configurations where one of the lines $p$, $q$ was rotated 
through 
$180^{\circ}$, as discussed in \cite{fnlrelns98}. This gives results for four cases where
the arguments of the function $\psi$ are themselves proportional to non-zero  powers 
of $x$, namely:

\ba \label{psi3}
 &&  \psi( x \alpha, x \beta ) \eq 1 + x \, [ (1-\alpha )/\beta - 2 ] + 
x^2  \, [ (1-\alpha) /\beta^2 + (-4+6 \alpha - \nonumber \\ 
 &&  \alpha^2)/\beta +  1/\alpha + 9 + 3 \alpha ] + 
x^3 \, [  (1-\alpha)/\beta^3 + (-5+7 \alpha - \alpha^2-\alpha^3)/\beta^2 +
\nonumber \\ && (2 /\alpha + 17 -    32 \alpha + 
 11 \alpha^2 - \alpha^3 )/\beta  -10/\alpha - 49 - 23 \alpha +2 \alpha^2 + 
(2/\alpha + 4 - \nonumber \\ && \alpha)\beta ] +  {\rm O } (x^4) \comma  \ea
\ba \label{psi4}
&& \ \psi ( x \alpha, \beta ) \eq 1 - 2 x + x^2 \, [(1-\alpha)/\beta +  1/\alpha +
9 + 3 \alpha +
(2/\alpha + 4 - \alpha ) \beta - \nonumber \\  &&
 (1-\alpha )^2 \beta^2/\alpha^2 ] + 
x^3 \, [  (-4+6\alpha-\alpha^2)/\beta -
10/\alpha - 49 -23 \alpha +2 \alpha^2 + 
\nonumber  \\  && 
(3/\alpha^2 -  19/\alpha-29+14\alpha-\alpha^2 ) \beta +
(-1/\alpha^3 + 13/\alpha^2-20/\alpha+11-
\nonumber \\   && 
3\alpha) \beta^2 -   2 (1-\alpha)^2 \beta^3/\alpha^3 ]
 +  {\rm O } (x^4)
\comma  \ea \ba \label{psi5}
&& \psi ( \alpha/x, \beta ) \eq 1 + \alpha + x \, [(1-\alpha)^2/\beta - 4 -4 \alpha + \beta ]
+ \nonumber \\
&&  x^2 \, [ (1-\alpha)(1-\alpha^2)/\beta^2 + (1/\alpha - 8 + 13 \alpha -
 8 \alpha^2 + \alpha^3 )/\beta -  2/\alpha + 25 +  \nonumber \\ &&   
 25 \alpha - 2 \alpha^2 + 
(1/\alpha - 5 + \alpha) \beta ]
 +   x^3 \, [   (1-\alpha) (1 - \alpha^3)/\beta^3 + (3/\alpha - 14 +
 \nonumber \\ && 
11 \alpha +  11 \alpha^2 - 14 \alpha^3 + 3 \alpha^4 )/\beta^2 +  
(1/\alpha^2-14/\alpha + 58 - 79 \alpha + 58 \alpha^2 - 
\nonumber \\
&& 14 \alpha^3 + \alpha^4 )/\beta  -  2/\alpha^2 +19/\alpha-146 - 146 \alpha + 
19 \alpha^2 -2 \alpha^3 +
\nonumber \\
&&  (1/\alpha^2-8/\alpha+30-8\alpha+\alpha^2) \beta ]
+ {\rm O } (x^4)
\comma  \ea \ba \label{psi6}
 && \psi ( \alpha /x , \beta /x ) \eq 1 + \alpha + \beta + 
x \, [-4 -4 \alpha + (1/\alpha-5+\alpha) \beta ] +  \nonumber \\ &&
x^2 \, [(1-\alpha)^2/\beta 
-2 /\alpha + 25 + 25 \alpha - 2 \alpha^2 + (1/\alpha^2 - 8 /\alpha + 
30 - 8 \alpha + ~~~  \nonumber \\ && \alpha^2) \beta ] 
 +   x^3 \, [ (1/\alpha-8+13 \alpha - 8 \alpha^2 + \alpha^3)/\beta - 
2/\alpha^2 + 19/\alpha - 146 -  \nonumber \\  && 
146 \alpha + 19 \alpha^2 -2 \alpha^3 + 
(1/\alpha^3 - 11/\alpha^2 + 56 /\alpha
-  175 + 56 \alpha - 11 \alpha^2 + \nonumber \\ && \alpha^3 ) \beta
+(1/\alpha^3-2/\alpha^2 + 2 /\alpha + 
2 -2 \alpha + \alpha^2) \beta^2 +  (-1/\alpha^3+3/\alpha^2-  \nonumber \\ && 
4/\alpha + 3 - \alpha ) \beta^3 ] + {\rm O } (x^4)
\period  \ea
In all of the expansions (\ref{psi1}) - (\ref{psi6}), the variables
$\alpha, \beta$ are of order one.

We can regard these expressions as giving the expansion of 
the function $\psi(\alpha, \beta )$ in various
domains in the $(\alpha, \beta)$ plane. We have verified that the results for neighbouring
domains are consistent with one another. For 
example (\ref{psi3}) and (\ref{psi4}) are consistent in that if one replaces 
$\beta$ in (\ref{psi4}) by $x \beta$, then
one obtains (\ref{psi3}), except for terms in (\ref{psi3}) that are of order 
higher than 3 when $\beta$ is of order $x^{-1}$. Similarly,
(\ref{psi1}) is consistent with both (\ref{psi4}) and (\ref{psi5}),
and (\ref{psi5}) is consistent with (\ref{psi6}). 

Taking $\alpha = {\rm O}(x^{-1})$
and  $\beta = {\rm O}(1)$ or ${\rm O}(x^{-1})$ in (\ref{psireln1}), one can readily 
check that (\ref{psi5})  and (\ref{psi6}) each has the consequent symmetry property.
Also, taking $\alpha = {\rm O}(x)$ and $\beta = {\rm O}(1)$ or
${\rm O}(x)$ in (\ref{psireln2}),  we find that the relation is satisfied by
(\ref{psi3})  and (\ref{psi4}).

If $q \rightarrow p$, the $p$ variables remaining finite, 
then $\alpha, \beta \rightarrow 1$
while $v$ and $\overline{v}$ remain finite. The coefficients of the terms proportional to
$v$ and $\overline{v}$ in (\ref{psi1})  then vanish, so we obtain the
unique result:
\be \psi(1,1) \eq 1 - x  + 7 x^2 -27 x^3 + {\rm O} (x^4) \period \ee
From (\ref{F1}) and (\ref{F2}) it follows that
$F_{pp}(1)/F_{pp}(0) = F_{pp}(2)/F_{pp}(0) = x \, \psi (1,1) $, so from (\ref{emmr})
\be \label{correct}
{\cal M}_1 \eq {\cal M}_2 \eq \frac{1 - F_{pp}(1)/F_{pp}(0)}{1 + 2 F_{pp}(1)/F_{pp}(0)} 
\eq 1 - 3 x + 9 x^2 - 45 x^3 + 231 x^4 + {\rm O}(x^5) \comma \ee
in agreement \cite{CTMII} with the conjecture (\ref{conj}).

\subsection*{Fourier transformed relations}

To obtain the relations (\ref{Foureqns}),  (\ref{Lrelns}) in terms of the variables 
$\alpha, \beta$ -- or rather in terms of $u, \beta$ -- 
define:
\be 
A(u, \beta) \eq 1  + x \beta^{-1} u^{-1} \, \psi (\alpha, \beta ) 
 + x \beta^{-1} u^{-2} \, \psi (\alpha, \alpha/\beta ) \comma \ee
\be
B(u, \beta ) \eq A(u, \beta) \, A(u, u^{-3} \beta^{-1}) \comma \ee
\be
\zeta (u,\beta) \eq \left[ 1 + u \beta \, \phi(\alpha) \, \phi(\beta )  +
 u^2 \beta \, \phi(\alpha) \, \phi(\alpha/ \beta )  \right] /\Delta (\alpha, \beta) \period \ee

Then $F_{pq}(0) = 1/A(u,\beta)$,
\be \label{FA}
\tilde{F}_{pq}(r) \eq  
     A(\omega^{-r} u, \beta )/ A( u, \beta )  \comma \ee
\be  \label{LB}
L_{pq} (r) \eq B(\omega^{-r} u, \beta )/B(\omega^{1-r} u, \beta ) \comma \ee
and, from (\ref{psireln1}) and (\ref{psireln2}),
\ba \label{frlnsA}
A(1/u, u^3 \beta )  \eq  A(1/u, x^{-1} \beta^{-1} ) & = &  A(u, \beta ) \comma \nonumber \\
 A(1/u, x/\beta ) \eq \zeta (u,\beta) \hspace{-4mm}
 && \hspace{-5mm} A(u,\beta) \comma \ea
\ba \label{frlnsB}
B(u^{-1}, u^3 \beta )  \eq &\!\!\!\! \! \! \!& B(u,u^{-3} \beta^{-1} ) \eq  B(u, \beta) \comma \nonumber \\
 B(u^{-1}, x \beta^{-1} ) &\!\!\!\! \! \! \!& \eq \zeta(u, \beta) \, B(u^{-1}, \beta^{-1} ) \period \ea

Together with (\ref{transab}) and the fact that $\Delta (\alpha, \beta) $ is a single valued function of $u^3$ 
and $\beta$, the relations (\ref{frlnsA}) imply
(\ref{Foureqns}), while (\ref{frlnsB}) imply (\ref{Lrelns}). One does not need the
definition  (\ref{defDelta}) of $\Delta  (\alpha, \beta) $.

Our function $\rho(z)$ is the $\tilde{\psi}(z)$ of \cite{hypfns98}. Another function discussed 
therein is
\be
h(\alpha, \beta ) \eq  \Phi(u, u^2 \beta) \,  \Phi(\omega u, \omega^2 u^2 \beta) \, 
 \Phi(\omega^2 u, \omega u^2 \beta) \period \ee
Using also eqns (32) and (58) of \cite{hypfns98}, it follows that
\be \label{defzeta}
\zeta (u,\beta) \eq  \tau (\alpha, \beta ) \, \Phi(u, u^2 \beta/x) / \Phi(u, u^2 \beta) \comma \ee
where \be
 \tau (\alpha, \beta ) \eq \frac{ h(\alpha, \beta ) }
{{Q(x)}^{2} \, {Q(x^3)}^{4} \, \rho(\alpha) \, \rho(\beta) \, 
\rho(\alpha/\beta ) \Delta  (\alpha, \beta)  }
 \period \ee

Obviously (\ref{FA}) and (\ref{LB}) are unaffected by multiplying the functions
$A(u, \beta)$, $B(u, \beta)$ by factors which are  single-valued functions of 
$u^3,\beta$ (i.e. of $\alpha, \beta$ ), such as  $h(\alpha, \beta)$ and 
$\tau ( \alpha, \beta )$.
To within such factors, using the symmetry property (\ref{Phisymm}), we see that  a simple solution
 of  (\ref{frlnsB}) is $B(u, \beta ) = \Phi(u,u^2 \beta ) $.
This implies that  $L_{pq} (r) = $ $
 \Phi(\omega^{-r} u, \omega^r u^2 \beta )/$ $\Phi(\omega^{1-r} u, \omega^{r-1} u^2 \beta )$, which is 
the spurious solution (\ref{spurious}) mentioned above.

More explicitly, let us define
\be 
C(u, \beta ) \eq \frac{B(u, \beta)} {\Phi(u,u^2 \beta )} \sep
D(u, \beta ) \eq \frac{C(u, \beta ) }{\left[ C(u, \beta ) \,
C(\omega u, \beta ) \, C(\omega^2 u, \beta ) \right] ^{1/3}} \period \ee
Thus $D(u, \beta )$ is $C(u, \beta )$ normalized by the multiplication of a 
function of $\alpha, \beta$ so that  $D(u, \beta ) D(\omega u, \beta ) D(\omega^2 u, \beta ) = 1$.
The equations (\ref{frlnsB}) then reduce precisely (including all factors) to the simple form
\be \label{eqnsD}
 D(1/u, u^3 \beta ) \eq  D(u,u^{-3} \beta^{-1} ) \eq D(u, \beta ) \eq D(u,  x \beta)  \period \ee

From our series expansion (\ref{psi1}), for 
$u, \alpha,  \beta$ of order 1, we find to order $x^4$ that
\ba \label{D1} & & \hspace{-8mm} D(u,1) = 1 - (u+u^{-1}) x + (1 \! + \! 2 u \! + \! 2 u^{-1}) x^2 + 
(u^3/3 + u^{-3}/3 \! - \! 11 u \! - \! 11 u^{-1} \! - \! 4) \, x^3 + \nonumber \\ & & \hspace{-5mm}  
(- 4 u^4/3 - 4 u^{-4}/3 -
 2 u^3 - 2 u^{-3} - 10 u^2/3 - 10 u^{-2}/3 + 56 u + 56 u^{-1} + 26) \, x^4  \ea
and
\ba \label{D2}
& & \hspace{-8mm} D(u, \beta ) /D(u,1)  =  1 + \left[ ( \alpha \!+\! 1 \!- \! 2 \beta) u +
( \alpha \! + \! 1 \! - \! 2 \alpha/ \beta)/u \right] \mu x^3 +
[ (2  \alpha  \beta+2  \beta-4  \alpha) u \mu^2 \nonumber \\ & & \hspace{-5mm}
+(2  \alpha^2/ \beta+2  \alpha/ \beta-4  \alpha) \mu^2/u +
(14  \beta - 8  \alpha - 8 + \beta^2 +  \beta^2/ \alpha) u \mu + 
(14 \alpha/ \beta - 8  \alpha - 8 + \nonumber \\ & &  \hspace{-5mm} \alpha^2/ \beta^2 +  \alpha/ \beta^2) \mu/u + 
(1- \beta)( \alpha- \beta) (u/ \beta^2 + 3 u/ \alpha +  \alpha^{-1}/ u + 3  \beta^{-2} /u ) 
\left. \right] x^4 \comma  \ea
while from (\ref{psi3}) - (\ref{psi6}), for $u, \alpha, \beta$ of orders $x^{1/3}, x, 1$,
 respectively, we find to order $x^4$ that
\ba & & \hspace{-8mm} 
D(u,1) \eq  1 - (u+u^{-1}) x + (1 \! + \! 2 u \! + \! 2 u^{-1}) x^2 + 
(u^3/3 + u^{-3}/3 \! - \! 11 u \! - \! 11 u^{-1} \! - \! 4) \, x^3 + \nonumber \\ & & \hspace{-5mm}  
(- 4 u^{-4}/3 -
  2 u^{-3}  - 10 u^{-2}/3 +  56 u^{-1} + 26) \, x^4 
  + (2 u^{-5} + 47 u^{-4}/3 + 58u^{-3}/3 ) x^5 - \nonumber \\ & & \hspace{-5mm}
 (3 u^{-7} + 7 u^{-6}/9) \, x^6 \ea
and
\ba  \label{D4}
& & \hspace{-8mm} D(u, \beta ) /D(u,1)  = 1 + (1-\beta ) (1- \beta x /\alpha ) \left\{
(u - 2 \beta u  + u^{-1} - 3 \alpha u^{-1} /\beta + 3 \alpha u^{-1}) x^3 + \right. \nonumber \\ 
& & \hspace{-5mm}
\left. (-\beta^2 u /\alpha - 2 \beta u /\alpha - 10 u^{-1} + 3 \beta u^{-1} - 3 u^{-1}/\beta) x^4 +
3 (1 + \beta + \beta^2) u^{-1} x^5 /\alpha \right\}\period \ea
Applying the  symmetries (\ref{eqnsD}) to these expansions, we can 
obtain $D(u,\beta)$ to order $x^4$ for 
$u$ of order $1$, $x^{1/3}$ and $x^{-1/3}$, and for all $\beta$. We have verified that the 
expansions are consistent with these symmetries, and the results for neighbouring domains
are consistent with one another in the sense discussed above, provided we remember that
some coefficients have poles when $a = 1, x^2, x^{-2}, \ldots $.

\subsection*{Order parameters in terms of $D(u, \beta )$}

From (\ref{emmr}) and the above equations we find that the order parameters ${\cal M}_1,
{\cal M}_2$ are given in terms of the function $D(u, \beta )$ by the simple relation
\be \label{MD}
{\cal{M}}_1^2 \eq {\cal{M}}_2^2 \eq  k^{2/3} \, \frac{D(\omega^{\pm 1}, 1)}{ D(1,1) }
\comma \ee
the factor $k^{2/3}$ being the contribution from the spurious result obtained by 
taking $D(u, \beta ) = 1$, i.e. by using (\ref{spurious}).  The right-hand side is 
to be evaluated in the limit $q \rightarrow p$, i.e. when the arguments $u, \beta$
of $D(u, \beta)$ behave so that  $u^3, \beta \rightarrow 1$, the ratio
$(\beta-1)/(u^3 \beta^2 - 1)$ remaining finite.

Using the expansion (\ref{D1}), we again obtain (\ref{correct}).

\section{Discussion}

If we can solve the functional relations for the ``split rapidity line''
correlation function $F_{pq}(a)$, then we can calculate the order parameters ${\cal M}_j$.
For the $N = 2$ chiral Potts model, i.e. the Ising model, this can be done 
straightforwardly, using a uniformizing elliptic function parametrization.
For $N \geq 3$ it is still an unsolved problem. For $N = 3$ we have written down the equations explicitly
using a hyperelliptic function parametrization: in (\ref{psireln1}) - (\ref{defDelta}) in
terms of the function $\psi(\alpha, \beta)$, 
and in (\ref{eqnsD}) in terms of $D(u, \beta)$. We have also presented the first 
few terms in a series expansion, obtained from finite lattice calculations.

The problem is how to solve the equations. One modest approach is to try to
guess the general form of the coefficients in the series expansions, and then see 
if the relations determine the coefficients precisely, at least successively term-by-term.
Such an approach works well for the Ising model in terms of its original variables
(the Boltzmann weights) \cite[eqns 2.15 to 2.30]{Enschede}. As a first step, we tried
replacing every numerical coefficient in (\ref{psi1}) -- (\ref{psi6}) by an arbitrary
number, substituted the expansions into (\ref{psireln1}) - (\ref{defDelta}), and attempted 
to systematically solve
the resulting equations for the unknown numbers. This works for $\psi(\alpha,\beta)$ to 
zero and first order in $x$, but 
at second order we were left with four undetermined coefficients, making it impossible to 
proceed further. (We did try looking ahead to see if the higher-order equations fixed these
four unknowns, but with no success.)

This approach appears to be even less successful when applied to the relations (\ref{eqnsD}):
for $D(u, \beta)$ to first order, which corresponds to zero order for 
$\psi(\alpha, \beta)$, we easily see that a solution is $D(u, \beta ) = 1 - \mu (u + u^{-1})
x$, where $\mu$ is an arbitrary constant. In fact it is painfully obvious that
the equations are unaffected by multiplying $D(u, \beta)$ by any function $f(u)$ 
satisfying $f(u) = f(u^{-1})$, $f(u) \, f(\omega u) \, f(\omega^2 u ) = 1$. This can certainly affect
${\cal M}_1$ and ${\cal M}_2$.

It could be very useful to merely know the functions to leading order for all values of
$\alpha, \beta$, or $u, \beta$: this would provide valuable clues as to the locations of any zeros 
or poles. This would be very significant if the functions are single-valued 
functions of their arguments, but we have no reason to suppose this, and know of no way of
obtaining even the leading-order behaviour except near the ``physical regime'' where the 
Boltzmann weights are real and positive (and  those regimes which map to it by the symmetries
used to derive the functional relations).

It may not be necessary to consider the full complex $u$ and $\beta$ planes. If we discard the first
equality in (\ref{eqnsD}), we  have two equations in which we can  regard $u$ as a fixed ``constant''.
They relate the values of $D$ with second argument $\beta$, $x \beta$ and $u^{-3}/\beta$, so 
are much like inversion relations \cite{Baxinv82}, the inversion points being 
$\beta = u^{-3/2}$ and $\beta = x^{1/2} u^{-3/2}$. If $D$ were analytic in an annulus in 
the complex $\beta$ plane, centre the origin, including these points, then it would be 
easy to solve
(\ref{eqnsD}) using a Laurent expansion in this annulus. In fact the solution would be a constant,
i.e. $D(u, \beta) $ would be independent of $\beta$.

To second order in $x$ this is the case, but from (\ref{D2}) we see that at third order
there are terms proportional to $\mu$, which from (\ref{defvvbw}) and (\ref{aub})
has a second-order pole at  $\beta = u^{-3/2}$. (At fourth order there are 
terms proportional to $\mu^2$.)
So $D$ is not analytic at the inversion points
and does depend on $\beta$.

We are left with a tantalizing puzzle: undoubtedly the functional 
relations do contain information, but they need to be supplemented with
a knowledge of the analyticity properties at the ``inversion point''. Perhaps
we should be using the individual hyperelliptic function variables \cite{Como93}
\be
z_p = x^{1/2} \, \exp [ 2 \pi i (s_p)_1] \sep w_p = \exp [ 2 \pi i (s_p)_2] \comma \ee
in terms of which $\alpha = z_q /z_p$, $\beta = w_q/w_p$. They can be chosen to be of
 order unity when $x$ is small, and to leading order
$w_p = z_p +1 = (\alpha-1)/(\alpha-\beta) $, 
$w_q = z_q +1 = \beta (\alpha-1)/(\alpha-\beta)$. Higher-order terms can be 
obtained from equations (4.5) and (4.6) of \cite{CTMII}.

A third variable
that enters the hyperelliptic parametrization of the Boltzmann weights is
$\gamma = w_p w_q /z_p$ \cite{hypfns98}. The variables $\alpha, \beta, \gamma$ all have 
the property that they are unchanged
by simultaneously replacing $p, q$ by $R^2 p, R^2 q$, so they automatically 
incorporate this symmetry of
the generalized correlation function. One can write the expansion in (\ref{psi1})
(at least to the order given) in a form where the coefficients are Laurent polynomials 
in $\alpha, \beta, \gamma$.  
 The trouble is that this expansion is not unique, since
$\alpha, \beta, \gamma$ are related to one another. Defining $\mu$ as in  (\ref{defvvbw}), 
i.e.
$\mu = \beta^{-1} (\beta - 1) ( \alpha - \beta)/(\alpha-1)^2$, we find  that
$$\mu \eq \gamma^{-1} + [\gamma/\alpha   - 3 - 3 \alpha^{-1} +(2 \alpha +
 2 \alpha^{-1}-1)/\gamma +
  (1+ \alpha)/\gamma^2] \, x  + {\rm O}
 (x^2) \period $$

Of course it may be that this hyperelliptic parametrization is not helpful at all, 
but this seems unduly pessimistic since (a) the parametrization does provide convenient 
and simple 
expressions for the  relevant Botzmann weight functions and their Fourier 
transforms \cite{hypfns98}, and
(b) the coefficients in the expansions are much simpler than they are if one uses 
the original variables $a_p, \ldots , d_p, a_q, \ldots , d_q$. While it is true that
the coefficients in our series expansions contain negative powers of $\alpha - 1$, there do not appear to be any 
negative powers of $\beta - 1$ or $\alpha - \beta$ (which occur in, say, the 
expansion of $\gamma$). It also appears from (\ref{D1}) that such negative powers 
disappear completely when $\beta = 1$, which is the case of interest for 
calculating the order parameters.

The above results are presented in the hope that they may be a step towards
verifying the conjecture (\ref{conj}) for 
the chiral Potts model order parameters.


\begin{thebibliography}{9}

\bibitem{fnlrelns98}
R.J.~Baxter,``Functional relations for the order parameters 
of the chiral Potts model'',  J. Stat. Phys. {\bf 91} (1998)
499 -- 524.

\bibitem{JMN93}
M.~Jimbo, T.~Miwa and A.~Nakayashiki, J. Phys. A {\bf 26} (1993)
2199 -- 2210.

\bibitem{Howes83} 
S. Howes, L.P. Kadanoff and
M. den Nijs, Nucl. Phys. B {\bf 215[FS7]} (1983) 169 -- 208.

\bibitem{AMPT89} 
G. Albertini, B.M. McCoy, J.H.H. Perk and S. Tang,
Nucl. Phys. B {\bf 314} (1989) 741 -- 763.

\bibitem{HenkelLacki89} 
M. Henkel and J. Lacki, Phys. Lett. A {\bf 138} (1989)
 105 -- 109.

\bibitem{Baxinv82}
R.J.~Baxter, J. Stat. Phys. {\bf 28} (1982) 1 -- 41.

\bibitem{Kyoto91}
R.J.~Baxter,  in {\it Proceedings of the International Congress of Mathematicians},
Vol. 2, I. Satake, ed. (Springer-Verlag, Berlin), (1991) 1305 -- 1317. 

\bibitem{CTMII}
R.J.~Baxter, J. Stat. Phys. {\bf 70} (1993) 535 -- 582.

\bibitem{Como93}
R.J.~Baxter, in {\it Integrable Field Theories}, eds. L. Bonora {\it et al},
Plenum Press, New York, (1993) 27 -- 37.

\bibitem{CTMP}
R.J.~Baxter, Int. J. Mod. Phys. B {\bf 7} (1993) 17 -- 28. 

\bibitem{hypfns98}

R.J.~Baxter, ``Some hyperelliptic function identities that occur
in the chiral Potts model'', to appear in J. Phys. A (1998).

\bibitem{RJB78}
R.J.~Baxter, Phil. Trans. Roy. Soc. {\bf 289} (1978) 315 -- 346.

\bibitem{BPauY88}
R.J.~Baxter, J.H.H.~Perk and H.~Au-Yang, Phys. Lett. A {\bf 128} (1988)
138 -- 142.

\bibitem{Enschede}
R.J.~Baxter, {\it in} ``Fundamental Problems in Statistical Mechanics V'',
ed. E.G.D. Cohen, North Holland (1980) 109 -- 141.

\end{thebibliography}
\end{document}